\documentclass[aps,preprint,nofootinbib,preprintnumbers,eqsecnum]{revtex4-1}

\usepackage[usenames, dvipsnames]{color}

\newcommand{\nb}[1]{\color{blue}}

\newcommand{\hl}[1]{\color{magenta}}

\usepackage[
	  pagebackref=false,
	  colorlinks=true,
      linkcolor=blue,
      urlcolor=blue,
      filecolor=black,
      citecolor=red,
      pdfstartview=FitV,
      pdftitle={},
        pdfauthor={},
        pdfsubject={},
        pdfkeywords={},
        pdfpagemode=None,
        bookmarksopen=true
      ]{hyperref}

\usepackage[normalem]{ulem}
\usepackage{amsmath}
\usepackage{enumerate}
\usepackage{amsfonts}
\usepackage{epsfig}
\usepackage{mathbbol}

\usepackage{setspace}

\newcommand\half{{\ensuremath{\frac{1}{2}}}}

\newcommand\field[1]{{\ensuremath{\mathbb{{#1}}}}}

\newcommand\vev[1]{{\ensuremath{\left\langle{#1}\right\rangle}}}

\newcommand\ket[1]{\ensuremath{\lvert{#1}\rangle}}

\newcommand{\RR}{\field{R}}

\newcommand{\be}{\begin{equation}}
\newcommand{\ee}{\end{equation}}
\newcommand{\bea}{\begin{eqnarray}}
\newcommand{\eea}{\end{eqnarray}}
\newcommand{\bega}{\begin{gather}}
\newcommand{\eega}{\end{gather}}

\newcommand{\bi}{\begin{itemize}}
\newcommand{\ei}{\end{itemize}}
\newcommand{\ben}{\begin{enumerate}}
\newcommand{\een}{\end{enumerate}}
\newcommand{\bca}{\begin{cases}}
\newcommand{\eca}{\end{cases}}
\newcommand{\bln}{\begin{align}}
\newcommand{\eln}{\end{align}}
\newcommand{\bst}{\begin{split}}
\newcommand{\est}{\end{split}}
\def\ie{\begin{equation}\begin{aligned}}
\def\fe{\end{aligned}\end{equation}}
\newcommand{\bma}{\le(\begin{matrix}}
\newcommand{\ema}{\end{matrix}\ri)}

\newcommand\al{{\alpha}}
\def\b{{\beta}}

\newcommand\lam{\lambda}
\newcommand\Lam{\Lambda}
\newcommand\om{\omega}
\newcommand\Om{\Omega}

\newcommand\ga{{\ensuremath{{\gamma}}}}
\newcommand\Ga{{\ensuremath{{\Gamma}}}}
\newcommand\de{{\ensuremath{{\delta}}}}
\newcommand\De{{\ensuremath{{\Delta}}}}

\newcommand\da{{\dagger}}

\newcommand\ov{\over}
\newcommand\ha{{\half}}

\def\le{\left}
\def\ri{\right}

\newcommand\sA{{\ensuremath{{\mathcal A}}}}

\newcommand\sF{{\ensuremath{{\mathcal F}}}}

\newcommand\sH{{\ensuremath{{\mathcal H}}}}

\newcommand\sL{{\ensuremath{{\mathcal L}}}}
\newcommand\sM{{\ensuremath{{\mathcal M}}}}
\newcommand\sN{{\ensuremath{{\mathcal N}}}}
\newcommand\sO{{\ensuremath{{\mathcal O}}}}
\newcommand\sP{{\ensuremath{{\mathcal P}}}}

\newcommand\sR{{\mathcal R}}
\newcommand\sS{{\mathcal S}}

\newcommand\vx{{\vec x}}
\newcommand\vk{{\vec k}}

\newcommand{\ktfd}{{\ket{\rm TFD}}}
\newcommand{\tfd}{{\rm TFD}}

\newcommand{\causal}{{causal connectability}}

\begin{document}

%\title{Emergence of Kruskal time and the black hole interior in holographic duality}

\title{Causal connectability between quantum systems and the black hole interior in holographic duality}

\preprint{MIT-CTP/5335}

\author{Samuel Leutheusser and Hong Liu}
\affiliation{Center for Theoretical Physics, 
Massachusetts
Institute of Technology, \\
77 Massachusetts Ave.,  Cambridge, MA 02139 }

\begin{abstract}

In holographic duality an eternal AdS black hole is described by two copies of the boundary CFT in the thermal field double state. This identification has many puzzles, including the boundary descriptions of the event horizons, the interiors of the black hole, and the  singularities. Compounding these mysteries is the fact that, while there is no interaction between the CFTs, observers from them can fall into the black hole and interact. We address these issues in this paper. 
In particular, we (i) present a boundary formulation of a class of in-falling {bulk} observers; (ii) present an argument that a sharp bulk event horizon can only emerge in the infinite $N$ limit of the boundary theory;  (iii) give an explicit construction in the boundary theory of an evolution operator for a bulk in-falling observer, making manifest the boundary emergence of the black hole horizons, the interiors, and the associated causal structure.  A by-product is a concept called causal connectability, which is a criterion for any two quantum systems (which do not {need to} have a known gravity dual) to have an emergent sharp horizon structure.

\end{abstract}

\today

\maketitle

%\tableofcontents

\bigskip
\bigskip

\section{Introduction}

\begin{figure}[h]
\begin{centering}
	\includegraphics[width=4.5in]{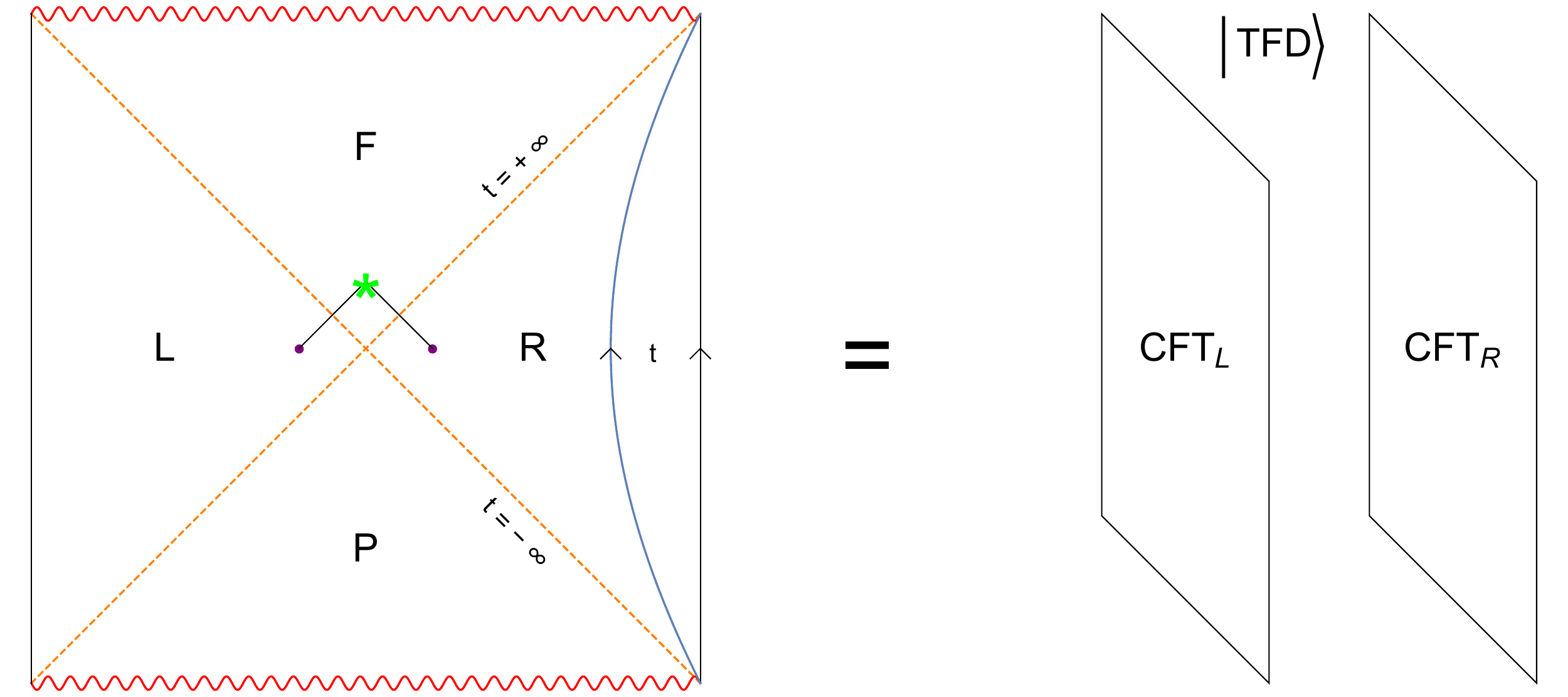}
\par\end{centering}
\caption{%\HL{The AdS-Rindler decomposition is shown on the left.} 
The Penrose diagram of an eternal black hole. The dashed lines are event horizons and the wavy lines are the singularities. 
Two observers from $R$ and $L$ can meet and interact behind the horizon despite the fact that there is no interaction between the left and right CFTs. 
}
\label{fig:casu}
\end{figure}

Understanding the emergence of causal structure in the bulk gravity theory from the boundary system 
in holographic duality has been an outstanding challenge. This issue can be formulated at many different levels. 
One of the most conspicuous features of bulk causal structure is the event horizon of a black hole. 
%presence of horizons, such as 
%the Rindler horizons of an AdS Rindler region and the event horizons of a black hole.
Understanding how these horizons and the regions beyond them emerge from the boundary theory should be a first step to address finer questions regarding bulk causal structure. 
 %The situation is particularly puzzling for the black hole case, which 
Consider an eternal black hole, which is dual to two copies of the boundary CFT in the thermal field double state $\ktfd$~\cite{Maldacena:2001kr}.  See Fig.~\ref{fig:casu}.
 The boundary time, $t$, for each copy of the CFT coincides with the bulk Schwarzschild time which ends at the horizon. How is it then possible to construct an explicit evolution operator in the CFT describing a bulk observer originally in region $R$ falling through the horizon into the $F$ region?\footnote{
 {\setstretch{1.0}
 In Jackiw-Teitelboim gravity it is possible to take an operator behind the horizon using symmetries, {as discussed in}~\cite{Maldacena:2018lmt,Lin:2019qwu}.
 There are also many different ways that boundary observables can probe regions behind the horizon see e.g.~\cite{Kraus:2002iv,Fidkowski:2003nf,Festuccia:2005pi,Hartman:2013qma,Liu:2013iza,Liu:2013qca,Susskind:2014rva,Grinberg:2020fdj,Zhao:2020gxq,Haehl:2021emt,Haehl:2021prg,Haehl:2021tft}, but in these discussions neither an emergent Kruskal-type time  nor the casual structure of the horizon was visible from the boundary. Similarly, ER$=$EPR type arguments~\cite{VanRaamsdonk:2010pw,Maldacena:2013xja} are largely concerned with a single time slice, not casual structure. 
See~\cite{Jafferis:2020ora} for an interesting recent discussion of in-falling observers using modular flows. Earlier discussions 
 of bulk reconstruction in the $F$ region include~\cite{Hamilton:2005ju,Hamilton:2006fh,Papadodimas:2012aq}.  
See also~\cite{Nomura:2018kia,Nomura:2019qps,Nomura:2019dlz,Langhoff:2020jqa,Nomura:2020ska}
for a description of the black hole interior from the perspective of coarse-graining.
}} How should we interpret the black hole singularities in the {bulk} $F$ and $P$ regions from the boundary theory?  Compounding these mysteries is the observation, first emphasized in~\cite{Marolf:2012xe}, that while there is no interaction between the two CFTs, observers from the $R$ and $L$ regions can fall into the $F$ region and interact with each other.

In this paper we address these questions. We first introduce a formulation of in-falling observers, which naturally leads 
to the concept of  casual connectability: a boundary {criterion} for an emergent {\it sharp} horizon in the dual gravity system. 
We then provide an explicit boundary construction of a one-parameter family of unitary operators, $U(s)$, that play the role of evolution operators along an in-falling trajectory. More explicitly, $U(s)$ has the following properties: 
\ben 

\item  It is generated by a Hermitian operator with a spectrum bounded from below.

\item  Consider a scalar field $\phi (X)$ in the $R$ region, i.e. with $X \in R$, and its evolution under $U(s)$: $\Phi (X;s) \equiv U^\da (s) \phi (X) U(s)$. {In this set-up,} there exists an $s_0 > 0$, such that for $s >  s_0$, $\Phi (X;s)$ can start having nonzero commutators with operators in CFT$_L$. Furthermore, $s_0$ coincides with the null Kruskal coordinate distance from $X$ to the horizon.

%along an in-falling trajectory can be described by

\item In the geometric optics limit (i.e. if the mass of $\phi$ is large), and with zero momentum along boundary spatial directions, $\Phi (X; s) = \phi (X_s)$ where $X_s$ is a bulk point. For $s < s_0$, $X_s \in R$, while for $s > s_0$, $X_s \in F$. 
\een

% We call the two copies of the CFT {\bf causally connectable} if there exists an evolution operator, $U(s)$, such that an evolved operator in CFT$_R$ can have nonzero commutators with operators in CFT$_L$. 

%

%Our starting point is the entanglement wedge reconstruction for the $R$ and $L$ regions of the eternal black hole, i.e. a local bulk operator $\phi (X)$  in the right region can be regarded as a boundary operator in CFT$_R$. 

The key to our discussion is the emergence, in the large $N$ limit of the boundary theory, of a type III$_1$ von Neumann algebraic structure from the type I boundary operator algebra and the half-sided modular translation structure associated to this type III$_1$ algebra. 
The black hole horizons, interiors, and singularities can all be understood as consequences of it. The type III$_1$ structure also sheds new light on the origin and nature of ultraviolet divergences in gravity. 
In this paper we outline the general ideas and the main results, leaving detailed expositions to~\cite{LL}.

\section{\causal: a boundary formulation of bulk horizon structure}

In this section we introduce a boundary formulation of a class of bulk in-falling observers and the associated signature of a sharp horizon.

%\footnote{
%See~\cite{} for other discussions of the puzzle. %In~\cite{} similar problem was considered in a tensor network model for an eternal %black hole. In the discussion there, the connected is used as an input rather than as an outcome. In~\cite{} the resolution lies in coupling the LR systems to other systems which generate coupling between L and R when integrating out other systems. But the discussion there will not address the second paradox above.
%} 

\subsection{A boundary formulation of in-falling observers}

In~\cite{Marolf:2012xe} a puzzle regarding the duality between the TFD state and the bulk eternal black hole geometry was raised. %The argument of~\cite{Marolf:2012xe} is very simple.  
Consider an initial state of the form %($A_L,$ is B_R$ are some Hermitian operators in the $L$ and $R$ CFTs, respectively)\footnote{\HL{should we just eliminate $B_R$?}}
\be \label{ehm}
\ket{\Psi_0} =e^{i A_L}  %e^{i B_R} 
\ktfd   %e^{i B_R}
\ee
where $A_L$ is a Hermitian operator in CFT$_L$ and we assume that its insertion only changes the energy of the system by an $O(1)$ amount such that its backreaction on the geometry can be neglected. %ake both $ e^{i A_L} $ and $e^{i B_R}$ to be
Since operators from the $R$ and $L$ sides commute, any measurement operator $M$ of the $R$ observer %at any time $t$ 
should commute with $e^{i A_L}$, i.e. 
\be\label{ehm1}
\vev{\Psi_0 |M |\Psi_0} =  \vev{\tfd |  M  |\tfd}  % \vev{\tfd | e^{-i B_R} M e^{i B_R} |\tfd} , 
\ee
so the presence of $e^{i A_L}$ cannot have any consequence on the measurement. But this appears to be in contradiction with the ability of the insertion of $e^{i A_L}$ to influence a right observer who has fallen into the $F$ region of the eternal black hole geometry, see Fig.~\ref{fig:casu}.

The above argument by itself does not directly pose a contradiction, as it 
assumes that the evolution of an in-falling observer from the $R$ region remains in CFT$_R$. 
It highlights, however, a seemingly counterintuitive requirement: for the identification of Fig.~\ref{fig:casu} to be correct, the description of an in-falling observer originally 
from the $R$ region {\it must} involve both the $R$ and $L$ systems. Indeed, from the causal structure of the black hole geometry, any operator in the $F$ region should involve degrees of freedom from both CFT$_R$ and CFT$_L$. Thus
 whatever measurement operator, $M$, the observer uses in the $F$ region must involve degrees of freedom in CFT$_L$, and we cannot assume that $M$ commutes with $e^{i A_L}$. 

%It may feel unintuitive that the evolution of an in-falling observer from the $R$ region must involve degrees of freedom of CFT$_L$, but there is no choice; this is forced on us if the duality is to hold.  (\HL{The above paragraph feels a bit repetitive. Can it be improved?}) 

%But this cannot be correct. From 
%In other words, the {\it evolution operator} for an in-falling observer must involve both the $R$ and $L$ systems,  
%From the usual story of bulk reconstruction~\cite{Hamilton:2005ju, Hamilton:2006az, Hamilton:2006fh}, an operator in the $F$ region should be expressed in terms of degrees of freedom of both CFT$_R$ and CFT$_L$. 
%It then appears that this issue is not an independent puzzle and simply reduces back to the question of how to describe the evolution of an in-falling observer using boundary language.  

%To sharpen the issue further, we now present an intrinsic boundary formulation of: (i) a class of bulk in-falling observers; (ii) {the causal structure due to a sharp horizon.} 

%\textcolor{red}{(What is the causal structure being associated to here? Is it the in-falling observers? Perhaps we can make this more clear?)} %causal structure associated with an event horizon, {\it intrinsically} in the boundary theory. 

%First consider the boundary formulation of in-falling observers. 

In this paper we will show that the evolution of a family of in-falling observers on the gravity side can be described by a {boundary} ``evolution operator'' $U(s)= e^{- i G s}, \, s \in \RR$ that satisfies the following properties: 

\ben

\item $G$ {involves} degrees of freedom from {\it both} CFT$_R$ and CFT$_L$. 

\item The Hermitian {generator} $G$ has a spectrum that is {\it bounded from below},
\be \label{soen}
G \geq 0 \ .
\ee

\een
%It appears natural to regard the above conditions as minimal requirements for a boundary formulation of an in-falling evolution. 
%Readers may wonder whether it is too restrictive for $G$ to be independent of $s$. 
The first property is needed for the in-falling evolution $\Phi (X;s) \equiv U (-s) \phi (X) U(s)$ of 
a bulk operator $\phi (X)$ with $X \in R$ to have support in the $F$ region. The second property is natural from the following perspectives: (i) if we interpret the eigenvalues of $G$ as energies for a family of bulk observers, they {should}  be bounded from below to ensure stability,\footnote{Note that $G$ should be understood as the ``energy'' associated with a full Cauchy slice in the black hole geometry rather than some local region. While some of the in-falling observers may only have a finite ``lifetime'' due to the presence of the singularity, they should nevertheless have a well-defined quantum mechanical description before hitting the singularity.} 
%even though due to the presence of singularities, 
 %is needed for the stability of the in-falling system, on the other hand 
(ii) the spectrum condition distinguishes $G$, as a generator of ``time'' flow,  from operators generating spacelike  displacements~(such as momentum operators). % As we will discuss later there are an infinite choices of such $U(s)$. 

%there must exist some $s_0 > 0$ such that the probability $p(s)$ must have the form 

\subsection{Sharp horizon structure only at infinite $N$} 

We will now show that the property~\eqref{soen} has important general implications regardless of the specific form of $U(s)$: a sharp event horizon can only emerge in the large $N$ limit of the boundary theory.

For this purpose, consider again the state~\eqref{ehm}, and the probability $p(s)$ {for an in-falling observer originally from the $R$ region to observe the existence of $e^{i A_L}$ along their ``trajectory'' parameterized by $s$.}  To reproduce the causal structure of the black hole spacetime, 
$p(s)$ should have the form
\be \label{ejn} 
p(s) = \bca 0 & s < s_0 \cr  \neq 0 & s > s_0 \eca , 
\ee
with $s_0 > 0$, as it is only possible to detect the influence of $e^{i A_L}$ after the horizon has been crossed. %\textcolor{red}{(Need to say anything about $e^{iA_L}$ supported near the horizon so horizon and its lightcone coincide?)}
The existence of such an $s_0$ and the non-smooth behavior of $p(s)$ at $s_0$ reflect the sharp causal structure from a sharp horizon.

There is a simple quantum mechanical argument~\cite{Hegerfeldt:1993qe} that the behavior~\eqref{ejn} is in fact not possible. 
%appears to be incompatible with general rules of quantum mechanics. 
Denote the projection operator that can detect the possible existence of $e^{i A_L}$ as $P_R$. The subscript $R$ emphasizes that this is an operator in CFT$_R$. The probability $p(s)$ can then be written as 
\be \label{enn1}
p(s) = \vev{\Psi_0| U^\da (s) P_R U (s)  |\Psi_0}  = \vev{\phi (s) |\phi (s)} , \quad \ket{\phi(s)} =  P_R e^{-i Gs} \ket{\Psi_0}   \ .
\ee
From~\eqref{soen}, we can analytically continue $U(s)$ to the lower half complex-$s$ plane. Accordingly, $\ket{\phi (s)}$ 
is a vector-valued analytic function of $s$ in the lower half complex $s$-plane, and is continuous along the real $s$-axis. %\textcolor{red}{(Is continuity obvious from the definition of $\ket{\phi(s)}$?)}
%in the lower half complex plane and so is $P_i (t)$. 
Equation~\eqref{ejn} means that 
$\ket{\phi (s)}$ vanishes for a finite segment, $s \in (0, s_0)$, of the real $s$-axis. Cauchy's theorem then says if $\ket{\phi (s)}$ is zero for any finite segment of $s$, it has to be identically zero for all $s$, incompatible with~\eqref{ejn}.  
Thus $p(s)$ can be zero only at isolated values of $s$ or identically zero, but cannot obey~\eqref{ejn}. 

This argument is very general, independent of details of specific states or quantum systems. {For example}, the two CFTs can interact and have a bulk geometry described by a traversable wormhole~\cite{Gao:2016bin}.

%: (i) the specific form of $\ket{\Psi_0}$ and thus applies to general entangled states between CFT$_{R, L}$, (ii) possible couplings between CFT$_{R, L}$ and thus also applies 

%but cannot 
%be zero for a finite interval, as in~\eqref{ejn}.  

%We now have a problem: the black hole geometry says that it is possible for observers of 
%CFT$_{R, L}$ to be causally connected by entering the $F$ region, but the general result from quantum mechanics says 
%this is not possible.
%We  refer to this as the puzzle of \causal.

%\sout{they are not} \textcolor{red}{left and right observers can never meet}.  

%\sout{causally connectable} \textcolor{red}{connected by a future region in the bulk}, 

\begin{figure}[h]
\begin{centering}
	\includegraphics[width=2.5in]{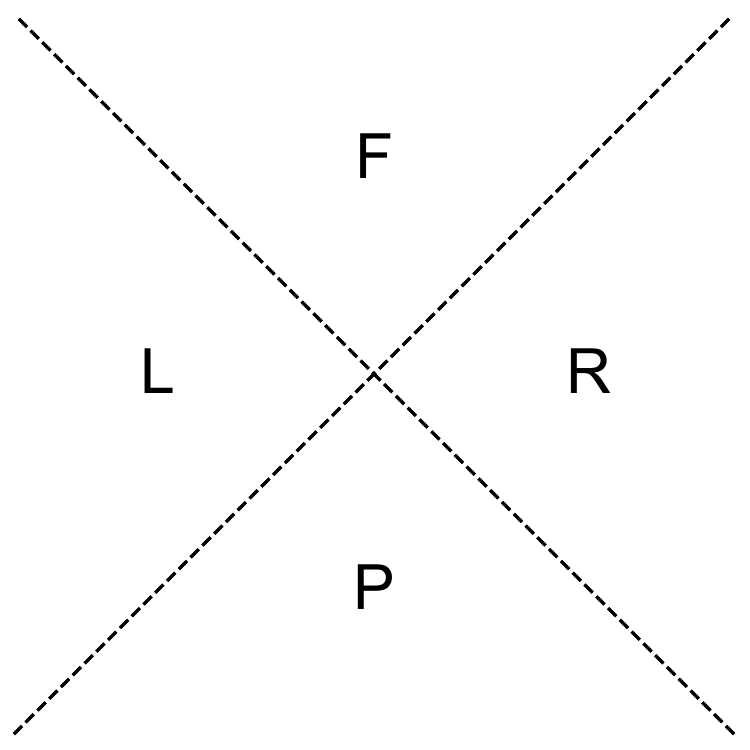}
\par\end{centering}
\caption{Rindler regions of Minkowski spacetime. 
}
\label{fig:rind}
\end{figure} 

Since the bulk gravity theory does have a sharp light cone, the above no-go argument must somehow be avoided 
in the duality relation. 
To understand {a possible resolution}, consider a closely related case: Rindler patches for a quantum field theory in Minkowski spacetime. See Fig.~\ref{fig:rind}. If we discretize the theory by putting it on a lattice,  the Minkowski vacuum $\ket{\Om}$ can be expressed as a TFD state for the $R$ and $L$ Rindler patches. In the discrete case there are no sharp light-cones. Any evolution on a lattice system has a small tail which gives rise to a nonzero commutator between two space-like separated operators. 
Indeed, in the discrete case, the above no-go argument applies:  observers from the $R$ and $L$ Rindler systems are either always connected (for $p(s)$ having only isolated zeros) or can never be connected (for $p(s)$ identically zero). 
However, in the continuum limit, they are separated by sharp light-cones, and can meet in the $F$ region only after evolution by some nonzero $s_0$. This difference in the sharpness of the light-cone structure between the discrete case and the continuum limit
can be attributed to a fundamental difference in the structure of their operator algebras. 
In the discrete case, the full Hilbert space factorizes into a tensor product of those of the $R$ and $L$ systems, and the operator algebras associated with the $R$ and $L$ systems are type I von Neumann algebras. 
 In the continuum limit, there is no local Hilbert space associated with the $R$ or $L$ patch, and the local operator algebra associated to a Rindler region is a type III$_1$ von Neumann algebra.\footnote{For reviews on the classification of von Neumann algebras see chapter III.2 of~\cite{Haag:1992hx} or section 6 of~\cite{Witten:2018zxz}.}
 In the continuum case, the no-go argument does not apply, since for a type III$_1$ von Neumann algebra, there does not exist any projector $P_R$ that can be used to detect the influence of an $L$ observer.\footnote{Any projector in a type III von Neumann algebra is infinite, and it is not possible to use such a projector to measure local excitations~\cite{Buchholz:1994eb,Yngvason:2014oia}.
Heuristically, due to the lack of a local Hilbert space associated with a Rindler region, there is no way to form finite projectors.} We expect that a type II von Neumann algebra will also be unable to describe operators outside of a sharp horizon but we will leave a rigorous mathematical proof elsewhere.

%In the discrete case, the discussion around~\eqref{enn1} applies and indeed a nonzero $s_0$ does not exist since, in a lattice system, operators in the $R$ and $L$ regions will have nonzero commutators for almost any $s$ (except possibly for isolated points). 

%and the $R$ and $L$ theories are causally connectable.
%a sharp horizon structure for the duality between the eternal black hole and two copies of a CFT in the TFD state.

The above Rindler story suggests a way to go around the no-go argument regarding~\eqref{ejn}. 
 The argument implicitly used that the full operator algebra of bounded operators of a CFT is type I (with the existence of a finite rank projector $P_R$), which is the case for the theory at finite $N$.\footnote{$N$ is the quantity that characterizes the number of boundary degrees of freedom, such as the rank of the gauge group or the central charge of the boundary CFT.} 
But the duality with the {\it classical} black hole geometry and the associated {\it sharp} causal structure needs to hold only in the large $N$ limit.  We will argue that in the $N \to \infty$ limit there is a pair of emergent type III$_1$ algebras, $\sM_{R, L}$, in the boundary theory.\footnote{Note that operator algebra associated with a local region in a QFT is type III$_1$. But here the emergent type III$_1$ algebras refer to those associated with the full boundary spacetime.}  The event horizons, black hole interior, and  singularities are all consequences of this emergence.

Given that the conditions~\eqref{ejn}--\eqref{enn1} for a sharp horizon structure cannot be defined for a type III$_1$ algebra,
we need a generalization.  We consider the function~\cite{Buchholz:1994eb,Yngvason:2014oia}
\be \label{hev1}
F(s) = \sup \le\{\le|\vev{\Psi_0|U^\da (s) \sO_R U(s) |\Psi_0} - \vev{{\rm TFD}|U^\da (s) \sO_R U(s) |{\rm TFD}}\ri| , \; \sO_R \in \sM_R , \; ||\sO_R|| \leq 1 \ri\} \ .
\ee
Existence of a sharp bulk horizon structure implies the existence of an $s_0 > 0$ and the behavior
\be \label{ejn1} 
F(s)= \bca 0 & s < s_0 \cr  \neq 0 & s > s_0 \eca 
\ .
\ee
For infinitesimal $A_L$, the above equation is the same as the existence of an $s_0 > 0$ and $\sO_R$ such that
\be \label{eja}
[A_L,  U^\da (s) \sO_R U(s)] \neq 0  \ , \quad s > s_0 \ .
\ee
 %Clearly the above definition applies to any state and is not specific to $\ktfd$. 

The condition~\eqref{ejn1} can be used to describe an emergent sharp horizon for any two quantum systems and general states, {even those without a known gravity dual}. We will refer to two systems in a state which satisfies~\eqref{ejn1}
as being causally connectable. %\textcolor{red}{(bold ``causally connectable''?)}

%Now we would like to argue that the sharp horizon structure defined by~\eqref{ejn}
%can {\it only} exist for the boundary theory at infinite $N$.

%Both $\sA_R$ and $\sA_L$ are cyclic and separating with respect to $\ktfd$, and they are commutant o

% and will demonstrate how they emerge using boundary theory language. 

\subsection{Emergent type III$_1$ von Neumann algebras} \label{sec:iii}

There is a natural candidate for the emergent type III$_1$ von Neumann algebra. %Suppose $\ktfd$ has inverse temperature $\b$. 
Consider the vector space of products of single-trace operators of CFT$_{R}.$ In the large $N$ limit, we can define an {\it algebra} of single-trace operators (see~\cite{Leutheusser:2022bgi} for details), $\sA_R,$ with respect to the thermal state, $\rho_{\beta}$ ($\b$ is the inverse temperature).\footnote{Strictly speaking, in the large $N$ limit the thermal state can only be defined through correlation functions obeying the KMS condition. No density matrix exists in $\sA_R,$ but we will continue to use the notation $\rho_\b,$ meant for the definition as a functional in the large $N$ limit.} We can build the GNS Hilbert space  $\sH^{\rm (GNS)}_{\rho_\b}$
of $\sA_R$ with respect to $\rho_\b$. We denote the representation of $\sA_R$ on $\sH^{\rm (GNS)}_{\rho_\b}$ as $\sM_R$. Here are some features of $\sH^{\rm (GNS)}_{\rho_\b}$: 

\ben 

\item The representation of $\rho_\b$ in $\sH^{\rm (GNS)}_{\rho_\b}$ is a pure state which we denote as $\ket{\Om_0}$. 
$\ket{\Om_0}$ is cyclic and separating for $\sM_R$. 

\item $\sH^{\rm (GNS)}_{\rho_\b}$ coincides with the GNS Hilbert space of the union of $\sA_L$ and $\sA_R$ with respect to 
$\ktfd$. In particular,  $\sM_L$, the commutant of $\sM_R$ in the operator algebra on $\sH^{\rm (GNS)}_{\rho_\b}$, can be viewed as 
 the representation of $\sA_L$ on $\sH^{\rm (GNS)}_{\rho_\b}$.\footnote{The emergence of this commutant algebra in $\sH^{(GNS)}_{\rho_\beta}$ is also discussed in appendix A of~\cite{Magan:2020iac}.}

%which is the representation of $\sA_L$, the algebra generated by single-trace operators of the CFT$_{L}$, on $\sH^{\rm (GNS)}_{\rho_\b}$. The commutant of $\sM_R$ is $\sM_L$.  (\HL{Is there a better way to say this?}) 

%$\sH^{\rm (GNS)}_{\rho_\b}$ coincides with the GNS Hilbert space of $\sA_R$ with respect to $U_L \ktfd$ where $U_L$ is an arbitrary unitary of CFT$_L$. \HL{In particular, they all have the same $\ket{\Om}$, but the corresponding $\sM_L$ are related by a unitary transformation?}  \HL{($U_L$ is not part of the algebra.)}

\een
We conjecture that $\sM_R$ and its commutant $\sM_L$ are type III$_1$ in the large $N$ limit.  An indication of this is the fact that the finite temperature spectral functions of single-trace operators have a continuous spectrum supported on the full real frequency axis despite the boundary CFT being defined on a compact space.\footnote{In~\cite{Festuccia:2005pi} the continuous spectrum has been argued to be responsible for the emergence of black hole horizons and singularities. See also~\cite{Festuccia:2006sa} for general arguments regarding the emergence of such a continuous spectrum.}

On the gravity side we quantize small metric and matter perturbations around the eternal black hole geometry. The resulting Fock space built on the Hartle-Hawking vacuum $\ket{HH}$ is denoted as $\sH_{\rm BH}^{\rm (Fock)}$ and the algebras of operators for the bulk theory in the $R$ and $L$ regions of the black hole are denoted as $\widetilde \sM_R, \widetilde \sM_L$. 
Under the duality we identify:
\be \label{ejnn}
\sH^{\rm (GNS)}_{\rho_\b} = \sH_{\rm BH}^{\rm (Fock)}, \quad
\ket{HH} = \ket{\Om_0} , \quad \sM_R = \widetilde \sM_R, \quad \sM_L = \widetilde \sM_L \ .
\ee
The above identifications essentially consist of the statement of bulk reconstruction for the $R$ and $L$ 
regions of the black hole. They should hold  perturbatively in the $1/N$ expansion~(or bulk $G_N$ expansion). 
That $\sM_{R, L}$ should be type III$_1$ is also required by the duality, as the bulk field algebras $\widetilde \sM_{R,L}$, being associated with local bulk regions in a quantum field theory in curved spacetime, must be type III$_1$.

 At the leading order in the $1/N$ expansion, the bulk algebras $\widetilde \sM_{R, L}$ are generated by a free field theory 
 while the boundary algebras $\sM_{R, L}$ are generated by a generalized free field theory (since representations of single-trace operators on $\sH^{\rm (GNS)}_{\rho_\b}$ are generalized free fields). 
 More explicitly, suppose a bulk field $\phi$ is dual to a boundary operator $\sO$.
 The restrictions $\phi_{R, L}$  of $\phi$ to the $R, L$ regions of the black hole are dual respectively to the representations in $\sH^{\rm (GNS)}_{\rho_\b} $ of $\sO_{R, L}$. They can be expanded in modes as 
   %then their  have the mode expansions 
   (the sums below should be viewed as a proxy for integrals)\footnote{In the equations below and the subsequent discussion $\sO$ should be viewed as the representation of a single-trace operator on the GNS Hilbert space.} 
\bega \label{bul}
\phi_R (t, \vx, w) = \sum_{\om , \vk} e^{- i \om t + i \vk \cdot \vx } v_{\om\vk} (w)   \, a_{\om \vk}^{(R)}   , \qquad
\phi_{L} (t, \vx, w) = \sum_{\om , \vk} e^{ i \om t - i \vk \cdot \vx }  v_{\om\vk} (w)  \, a_{\om, \vk}^{(L)} ,   \\
\sO_{R} (t, \vx) = \sum_{\om , \vk} e^{- i \om t + i \vk \cdot \vx } N_{\om \vk}  \, a_{\om \vk}^{(R)}   , \qquad
\sO_{L} (t, \vx) = \sum_{\om , \vk} e^{ i \om t - i \vk \cdot \vx } N_{\om \vk} \, a_{\om, \vk}^{(L)} ,
\label{bdt}
\end{gather} 
where $\vx$ collectively denotes the spatial coordinates on the boundary and $w$ denotes the bulk radial coordinate. 
$v_{\om \vk} (w)$ are the bulk mode functions and the $N_{\om \vk}$ are constants. The identifications~\eqref{ejnn} are reflected in the fact that $\phi$ and $\sO$ share the same creation/annihilation operators $a_{\om \vk}^{(R,L)}$, and thus $\phi_{R,L}$ can be viewed as 
elements of boundary algebras $\sM_{R, L}$. 

\begin{figure}[h]
\begin{centering}
	\includegraphics[width=2.5in]{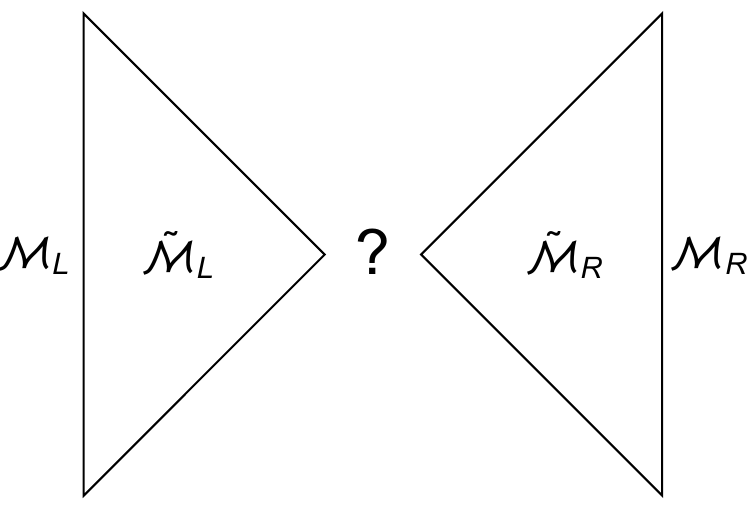} 
\par\end{centering}
\caption{The identifications~\eqref{ejnn} establish duality of CFT$_{R, L}$ with the exteriors of the black hole spacetime, but they do not directly say anything about the existence of the $F, P$ regions of a connected bulk.
}
\label{fig:doleftRightConnect}
\end{figure}

The identifications~\eqref{ejnn} are shown in Fig.~\ref{fig:doleftRightConnect}.  In subsequent sections we will show that the type III$_1$ nature of $\sM_{R, L}$  leads to the emergence of the $F$ and $P$ regions and the associated causal structure. 

%At this stage

\section{Emergent new times in the boundary} 

In this section we discuss how to generate  new times in the boundary theory. 
Our main tool is  half-sided modular translation~\cite{Borchers:1991xk, Wiesbrock:1992mg}, and an extension of it. 
Suppose $\sM$ is a  von Neumann algebra and the vector $\Om$ is cyclic and separating for $\sM$. 
We denote the corresponding  modular flow and conjugation operators as $\De_\sM$ and $J_\sM$. 
% and its action on an element of $\sM$ has the form 
%\be 
%\De_\sM^{-it} A \De_{\sM}^{it} = e^{i K_\sM t} A e^{- i K_\sM t}, \quad A \in \sM , \quad K_\sM \equiv- \log \De_\sM  \ .
%\ee
Now suppose there exists a  von Neumann subalgebra $\sN$ of $\sM$ with the properties:

\ben 
\item $\Om$ is cyclic for $\sN$ (it is automatically separating for $\sN$ as $\sN \subset \sM$). 

\item The half-sided modular flow of $\sN$ under $\De_\sM$ lies within $\sN$, i.e. 
\be \label{ghb}
\De^{-it}_\sM \sN \De^{it}_\sM   \subset \sN , \quad t \leq 0 \ .
\ee
\een
It can then be shown~\cite{Borchers:1991xk, Wiesbrock:1992mg,Borchers:1998ye}  that there exists a unitary group $U(s), \, s \in \RR$ with the following properties:

\ben 

\item $U(s)$ has a positive generator, i.e. 
\be\label{enn}
U(s) = e^{-i G s}, \qquad G \geq 0
\ee

\item It leaves $\Om$ invariant
\be 
U(s) \Om = \Om, \quad \forall s \in \RR
\ee

\item Half-sided inclusion %\textcolor{red}{This should have $\sM \rightarrow \sN$ below? (BY (2.8)) Both are true (also BY (2.47))}
\be \label{hsi}
U^\da (s) \sM U (s) \subseteq \sM , \quad \forall s \leq 0 \ .
\ee

\item $\sN$ can be obtained from $\sM$ with an action of $U$
\be \label{mo}
\sN = U^\da (-1) \sM U(-1) 
\ee
and with $\sN_t \equiv \Delta_{\sM}^{-it} \sN \Delta_{\sM}^{it}$
\be \label{bdrytshift}
e^{i G s} \sN_t e^{- i G s}  = \sN_{f_1 (s,t)}, \quad f_1 (s, t) = - {1 \ov 2 \pi} \log ( e^{-2 \pi t} - s) \ .
 \ee

\een

\iffalse
 \textcolor{red}{For the following statements we need to understand how our generator $G$ relates the generator $G_{BY} \equiv \log\Delta_\sN - \log\Delta_\sM$ discussed in the BY paper, including the relative normalization. From our studies on the region of support of an evolved boundary operator we learned that our $U(s)^\da$ acts in the exact same manner as BY's $\Gamma(\tau)$ when we identify $\tau = \frac{\beta s}{2\pi}$. Since $\Gamma\left(\frac{\beta s}{2\pi}\right) = \exp\left( \frac{i G_{BY}}{\beta} \frac{\beta s}{2\pi} \right) = \exp\left( i \frac{G_{BY}}{2\pi} s\right) = \exp\left(i G s\right) = U(s)^\da$ we identify $G = \frac{G_{BY}}{2\pi}$. We see then that $\Gamma\left(-\frac{\beta}{2\pi}\right) = U(-1)^\da$ confirming the equation below as equivalent to BY's (2.48) for ``minus'' half-sided modular translations}
\textcolor{red}{The definition we should give is $\sN_t \equiv \Delta_{\sM}^{-it} \sN \Delta_{\sM}^{it}$ so that we do not need to introduce the boundary Hamiltonian. With this definition of the time-translations of $\sN$ and the identification $G = \frac{G_{BY}}{2\pi}$ BY (3.10) implies the equation above, confirming the form of $f_1$.}
\fi

We also note that 
\be \label{uun3}
%\De^{-it}_\sM U(s) \De^{it}_\sM = U (e^{2 \pi t} s), \qquad  
J_\sM U(s) J_\sM = U(-s) 
\ee 
and since $\sM' = J_\sM \sM J_\sM$ 
\be 
U^\da(s) \sM' U (s) \subseteq \sM' , \quad s \geq 0 \ . 
\ee
The above structure is called half-sided modular translation and exists only if $\sM$ is a type III$_1$ von Neumann algebra~\cite{Borchers:1998}.

%\textcolor{red}{This is also true if we conjugate by $\De_{\sN}$? (BY (2.9)) but this $\De_\sM$ is more relevant for the later discussion? Note that below I changed the RHS factor from $e^{-2 \pi t}$ to $e^{2 \pi t}$ as I believe this is what follows from BY (2.28)}

In our context, we take $\sM = \sM_R$, $\sM' = \sM_L$, and $\Om = \Om_0$ (introduced in Sec.~\ref{sec:iii}), with the modular operator $\De_\sM =\exp \le(\b ( H_R - H_L)\ri)$
where $H_{R, L}$ are, respectively, the Hamiltonians of CFT$_{R,L}$. Thus modular evolutions of $\sM_R$ under $\De_\sM$ are simply 
the standard boundary time translations. 
By choosing different $\sN$ we can construct different one-parameter group evolutions. 
These are candidates for new emergent ``times'' as the corresponding generators are bounded from below, as in~\eqref{enn}.
At leading order in the $1/N$ expansion,  $\sM_R$ and $\sM_L$ are generated by generalized free fields. With the mode expansion~\eqref{bdt}, using~\eqref{hsi} the action of $U(s)$ on $\sO_{R}$ can be written in terms of a linear transform on $a_{\om \vk}^{(R)}$, 
\be \label{ena}
U^\da (s) a_{k}^{(R)}  U(s) = \sum_{k'} C_{k k'} (s) a_{k'}^{(R)} , \quad k = (\om, \vk)  , \quad s \leq 0 \ .
\ee
When $s > 0$, $U(s)$ takes $\sO_{R}$ outside $\sM_R$ and the evolved operator is no longer covered by the theorem. 

It turns out that when $\sM_R$ is generated by generalized free fields, much more can be learned and it is possible to generalize the action of $U(s)$ to all values of $s$~\cite{LL}:

\ben 

\item The matrix $C_{kk'}(s)$ can be determined up to a phase 
\be \label{genSt} 
C_{kk'} (s) =  \de_{\vk, \vk'} \sqrt{{\sinh \pi |\om'| \ov \sinh \pi |\om|}} e^{i \ga_{\om}  (\vk) - i \ga_{\om'}  (\vk)} 
(-s)^{- i (\om -\om')} \Ga (i (\om-\om') ) 
\ee
where the phase $e^{i\ga_\om (\vk)}$ depends on the specific system and the choice of $\sN$.

\item For all $s$, we can write 
\be \label{hdn}
U^\da (s) a_{k}^{(\al)}  U(s) = \sum_{k'} \sum_{\b =R, L} \Lam_{k k'}^{\al \b} (s) a_{k'}^{(\b)} , \quad \al = R, L  \ .
\ee
For $\al =R$ and $s < 0$, equation~\eqref{hdn} reduces back to~\eqref{ena} which means that 
\be
\Lam_{kk'}^{RR} (s) = C_{kk'} (s),  \quad  \Lam_{kk'}^{RL} (s) = 0, 
\quad s < 0 \ .
\ee
 $\Lam^{R \b}_{k k'} (s)$ for $s> 0$ can also be expressed in terms of $C_{kk'} (s)$, 
\be \label{jen}
\Lam_{kk'}^{RR} (s) = {\sinh \pi \om \ov \sinh \pi \om'}   C_{kk'} (-s), 
\quad \Lam_{kk'}^{RL} (s) = {\sinh \pi (\om + \om') \ov \sinh \pi \om'}   C_{k, -k'} (-s), \quad 
 s > 0  \  .
\ee
The action of $U(s)$ on $\sO_L$, i.e. $\Lam^{\al \b}$ with $\al =L$, can be obtained from the relation 
\be 
\Lam_{kk'}^{\al\b} (s) = \Lam_{-k ,-k'}^{\bar \al \bar \b} (-s) 
\ee
which is a consequence of~\eqref{uun3} and $\sM' = J_\sM \sM J_\sM$.  

\een 
%\HL{To conclude we note that the unitary operators, $U(s)$, are not elements of $\sM_R$ or $\sM_L$, and should be viewed as automorphisms of the union of these algebras.}

\iffalse
In this case, we can write the action of $U(s)$ on $\phi^{(\al)}$ more explicitly. 
With generalized free fields $\phi^{(\al)} (t, \vx)$ expanded in Fourier modes 
\be 
\phi^{(R)} (t, \vx) = \sum_{\om , \vk} e^{- i \om t + i \vk \cdot \vx } N (\om , \vk)  \, a_{\om \vk}^{(R)}   , \qquad
\phi^{(L)} (t, \vx) = \sum_{\om , \vk} e^{ i \om t - i \vk \cdot \vx } N (\om , \vk) \, a_{\om, \vk}^{(L)}    , 
\ee
from~\eqref{hsi} the action of $U(s)$ on $\phi^{(R)}$ for $s < 0$ should reduce to a linear transformation on $a_{\om \vk}^{(R)}$. 
When $s > 0$, the action of $U(s)$ should take $\phi^{(R)}$ outside of $\sM_R$. In other words, the transformation of $a_{\om \vk}^{(R)}$ must now involve $a_{\om \vk}^{(L)}$. 
\fi

\section{Emergence of the bulk Rindler horizon from the boundary} \label{sec:rind}

\begin{figure}[h]
\begin{centering}
	\includegraphics[width=2in]{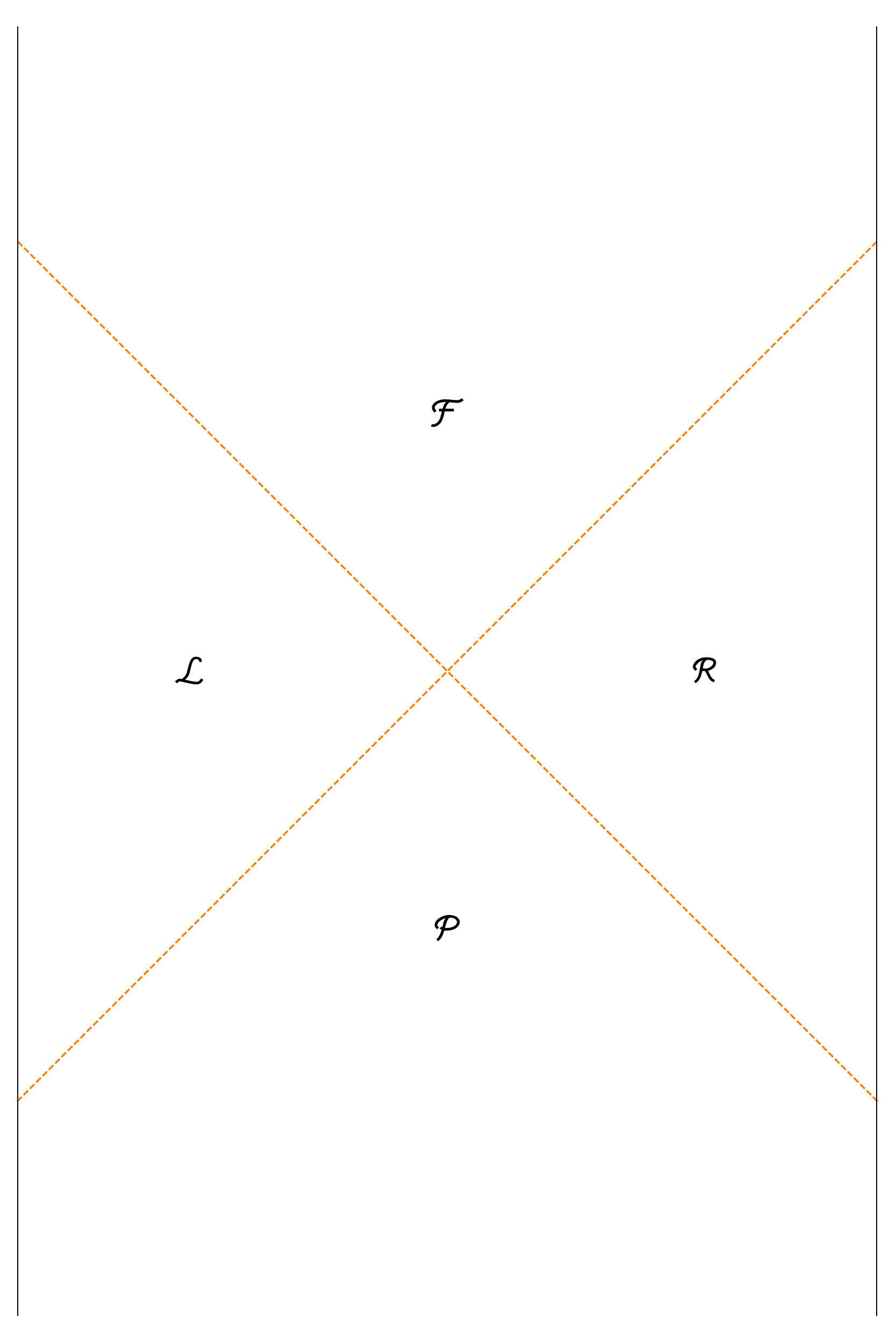} \qquad 
\par\end{centering}
\caption{AdS Rindler regions of the bulk spacetime. The vertical lines denote the boundary and the dashed lines are Rindler horizons. 
}
\label{fig:Arind}
\end{figure}

As a warmup for the black hole story we consider the emergence of the bulk Rindler horizon from the boundary system using the method outlined in the last section. Our input is the duality between the AdS Rindler region and the boundary CFT in the corresponding Rindler patch~\cite{Hamilton:2006az,Czech:2012be, Morrison:2014jha}.
%We construct  evolution $U(s)$ operator which can take a local operator in the $R$ region along an in-falling trajectory through the Rindler horizon. 
We will consider as an illustration, a bulk theory in the Poincare patch of AdS$_3$,  dual to a two-dimensional boundary CFT on $\RR^{1,1}$. 
The bulk spacetime contains two AdS Rindler regions which respectively have the two Rindler regions in $\RR^{1,1}$ as their boundaries. The bulk theory in the bulk $\sR/ \sL$ region in Fig.~\ref{fig:Arind} is ``reconstructible'' from the boundary theory in the corresponding boundary $R/ L$ region in Fig.~\ref{fig:rind}. In this case, $\sM_{R/ L}$ is the algebra generated by single-trace operators in the $R/L$ Rindler regions.  We {note that} going beyond the AdS Rindler horizon can be achieved by symmetries\footnote{See~\cite{Magan:2020iac} for a discussion. Going behind the horizon of a black hole in Jackiw-Teitelboim gravity~\cite{Maldacena:2018lmt,Lin:2019qwu} is also similar to the AdS Rindler case, {as it can be done using a symmetry operator.}}
 as the usual AdS isometries (which are dual to boundary conformal symmetries) can take an operator in the $\sR$ region to the $\sF$ region. %without using the half-sided modular translation structure. \textcolor{red}{(Do we need to say `without using half-sided modular translation'? The operator one gets should be the same but there is no need to invoke half-sided modular translations to construct it)} 
 But this example provides a nice illustration of the method of the last section and an interesting contrast for the discussion of the black hole case where such symmetries do not exist.

More explicitly, the metric for an AdS Rindler region can be written as 
\be \label{Rindc}
    ds^2 =  \frac{R^2}{w^2}\left[-\left(1 - w^2  \right)d\eta^2 + \left(1 - w^2  \right)^{-1}dw^2 + d\chi^2  \right] 
    \ee
with the Rindler horizon at $w =1$ and boundary at $w=0$. Consider a bulk scalar field $\phi$ dual to a boundary operator $\sO$ with dimension $\De$. The restriction $\phi_R (X)$ of $\phi$ to the $\sR$ region (with $X = (\eta, w, \chi)$) can be expanded in modes as 
\bega \label{ebs}
 \phi_R (X) = \int \frac{d^2 k}{(2\pi)^2} \,   e^{i k \cdot x } v_{k} (w) a_{k}^{(R)}  , \quad k = (\om, q), \quad 
 k \cdot x = -  \om \eta + q \chi \\
%\ee
%where 
%\bega
\label{qdef}
v_k  (w) = N_{k}  f_{k} (w) , \quad  q_\pm = \ha (\De + i (\om \pm q)), \quad  \bar q_\pm = \ha (\De - i (\om \pm q)) \\
N_k %= {1 \ov \sqrt{2} \Ga (\De) } {|\Ga (q^+)| |\Ga (q^-)| \ov \sqrt{|\om|} |\Ga (i \om)|}
 = {\sqrt{\sinh \pi |\om| }\ov \sqrt{2 \pi} \Ga (\De)} \le|\Gamma\left(q_+ \right) \Gamma\left(q_- \right) \ri| , \quad
   f_{k} (w) =  w^{\Delta} (1-w^2)^{-i\omega/2} {}_2F_1\left(\bar q_-  , \bar q_+ ; \Delta;  w^2\right)
    \ .
\end{gather} 
%where the explicit form of $v_k (w)$ is given in Appendix~\ref{app:mode}. 
%are a set of mode functions depending only on the radial coordinate $w$. 
The $a_k^{(R)}$ are creation {(for $\omega < 0$)} and annihilation {(for $\omega > 0$)} operators of the boundary generalized free field theory in the $R$ region, and thus 
$\phi_R (X)$ can be interpreted as an operator in the boundary theory. 
There is a similar ``bulk reconstruction'' equation for $\phi_L$  in terms of $a_k^{(L)}$. 

\begin{figure}[h]
\begin{centering}
	\includegraphics[width=2.5in]{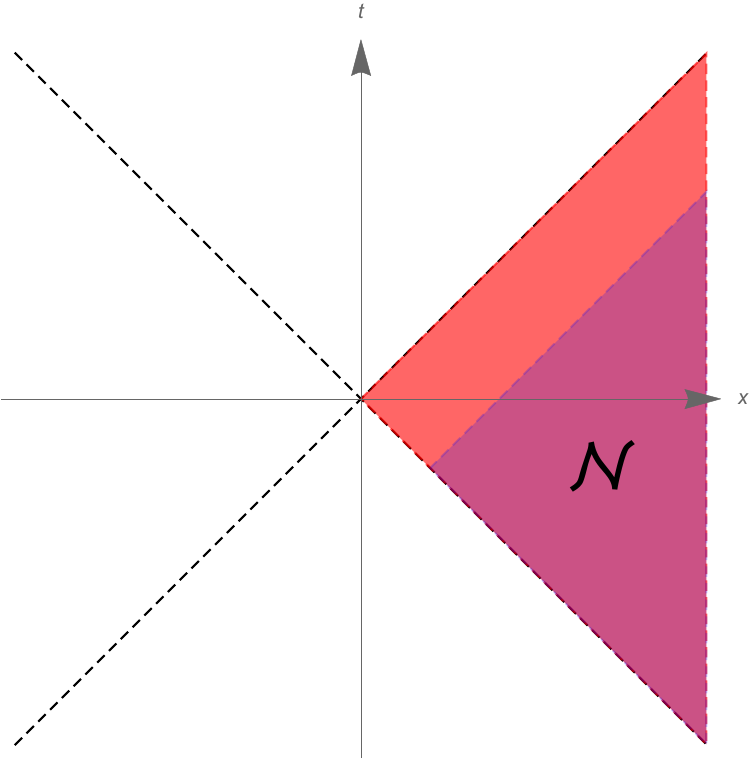} \qquad  \includegraphics[width=2.5in]{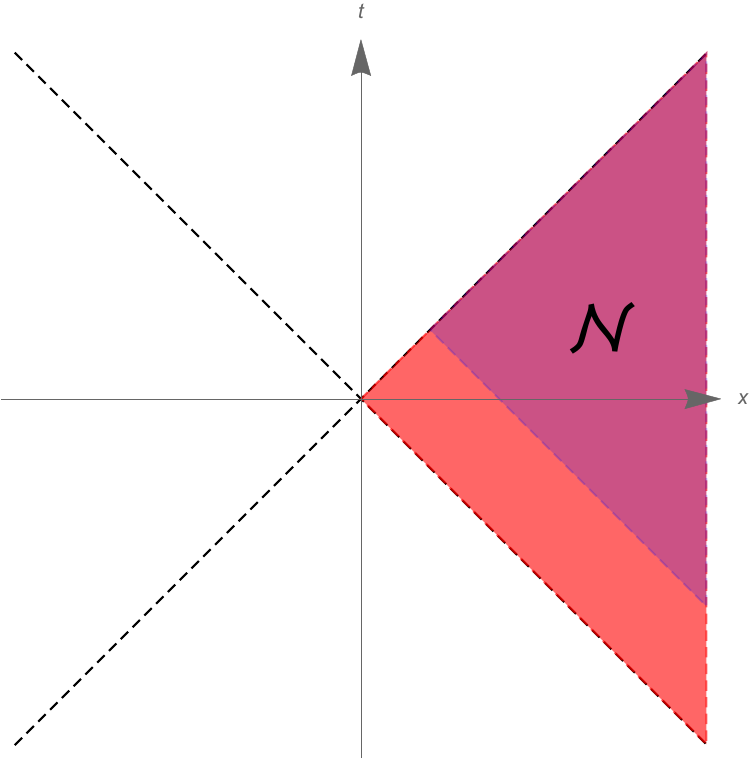}
\par\end{centering}
\caption{Boundary subregions that are used for the construction of a one-parameter group of unitaries implementing bulk Poincar\'e $U-$ and $V-$translations, respectively.  $t, x$ are boundary Minkowski coordinates. 
}
\label{fig:shiftedBdryRindlerWedge}
\end{figure} 

Our goal is to use the boundary theory in the $R, L$ regions and~\eqref{ebs} to reconstruct the bulk theory in the full Poincare AdS spacetime, including the $\sF$ and $\sP$ regions in Fig.~\ref{fig:Arind}. 
For this purpose we take $\sM = \sM_R$, which is type III$_1$ as it is the local algebra for a Rindler region. We take $\sN$  
to be the algebra of operators in the region indicated in Fig.~\ref{fig:shiftedBdryRindlerWedge}, corresponding to a null shift in one of the boundary light cone directions. Clearly $\sN \subset \sM$. 
For the choice of the left plot of Fig.~\ref{fig:shiftedBdryRindlerWedge}, the corresponding $C_{k k'}$ can be computed explicitly in the boundary theory. 
It has the form of~\eqref{genSt} with 
\be \label{khs}
e^{i \ga_k} ={\Ga (\bar q_+) \ov |\Ga (\bar q_+)|}  \ .
\ee
From our earlier discussion this fully determines $\Lam_{kk'}^{\al \b} (s)$ for all $s$. We can then work out how a bulk field~\eqref{ebs} transforms under the evolution of $U(s)$. From the form of $v_k (w)$ %in Appendix~\ref{app:mode}, 
it can be shown that 
for $s < s_0 \equiv  e^{- \eta + \chi} \sqrt{1-w^2}$ 
\be \label{ehen}
U^\da (s) \phi_R (X) U(s) = \phi_R (X_s),
\ee
where $X_s = (\eta_s, \chi_s, w_s)$ with 
\be\label{nsh}
w_s = {w \ov \sqrt{1-a_s}} , \quad e^{\eta_s} = {e^{\eta} \sqrt{1-w^2} \ov \sqrt{1 - a_s - w^2}}, 
\quad e^{\chi_s} = e^{\chi} \sqrt{1 - a_s} , \quad 
a_s \equiv  {s (1-w^2) \ov s_0} \ . % u e^{\xi^-} \sqrt{1-w^2} 
%e^{\xi^-_u} = {e^{\xi^-} \sqrt{1-w^2} \ov \sqrt{1 - a_u}\sqrt{1 - a_u - w^2}} ,
%\quad  e^{\xi^+_u} = {e^{\xi^+} \sqrt{1-w^2} \sqrt{1 - a_u} \ov \sqrt{1 - a_u - w^2}}  \ .
\ee
It can be readily checked that the above transformation precisely corresponds to a shift $U \to U + s$ of the Poincare null coordinate $U$, in terms of which the AdS metric reads\footnote{The two sets of coordinates are related 
by $t = R e^{\chi} \sqrt{1-w^2} \sinh\eta, \; x = R e^{\chi} \sqrt{1-w^2} \cosh\eta, \; z = R e^{\chi} w$.}
\be \label{poiC}
ds^2 = {R^2 \ov z^2} \le(- dt^2 + dx^2 + dz^2 \ri) = {R^2 \ov z^2} \le(- dU dV + dz^2 \ri), \quad U = t-x , \quad V = t+ x \ .
\ee
In particular, $-s_0$ is the value of the $U$ coordinate for the point $X$, and thus $s = s_0$ takes $X$ to the Rindler horizon. The transformation~\eqref{ehen} is smooth at $s =s_0$, and when $s > s_0$ 
the right hand side also involves $a_k^{(L)}$ with
\be 
U^\da (s) \phi_R (X) U(s) %= \phi_R (X_s) + \phi_L (X_s) 
= \phi_F (X_s),
\ee
where $X_s \in \sF$ is again obtained by a null Poincare shift $U \to U  + s$ 
and $\phi_F (X_s)$ is the restriction of $\phi$ to the $\sF$ region. 

%and lies in the future region. 
%\HL{In the above equation $\phi_R, \phi_L$ are a certain analytic continuation of $\phi_{R, L}$ to the future region and $\phi_F $ is the full expression for $\phi$ in the future region $\sF$.} \textcolor{red}{(This seems clear to me but maybe not to someone less familiar?)}

%i.e. the evolved operator is precisely given by the expression for $\phi$ in the future region $\sF$ with $X_u$ again determined .  

Similarly, choosing $\sN$ as in the right plot of Fig.~\ref{fig:shiftedBdryRindlerWedge} gives rise to null Poincare translations $V \to V + s$. 

The above construction explicitly demonstrates the emergence of the AdS Rindler horizons and the associated bulk causal structure from the boundary theory. 

%unitary evolutions $V(v)$ which correspond to . $U(u)$ and $V(v)$ can then be used to reconstruct the full AdS spacetime. 

%which precisely recovers the bulk reconstruction of $\phi$ in the future region. 
%This also reconstructs the causal structure of the bulk theory. 

% the above duality between subregions, construct a {\it boundary} evolution operator $U(s)$ which when acting on $\phi_R$ can take it to beyond the $\sR$ region and obtain the reconstruction formula for a bulk field the full AdS spacetime. 

%The bulk operator algebras are dual to the boundary operator algebra in the Rindler region. 

\section{Emergence of the black hole event horizon and Kruskal time} \label{sec:btz}

We now consider the emergence of the interior of the black hole from the boundary theory.
The bulk theory in the eternal black hole geometry is now dual to two copies of the boundary theory on $\RR \times S^{d-1}$ in the state $\ktfd$. For illustration we again take $d=2$, i.e. a BTZ black hole~\cite{Banados:1992wn}, whose metric has exactly the same 
form as~\eqref{Rindc} but now with $\chi$ compact.\footnote{In units of~\eqref{Rindc}, the inverse Hawking temperature is $\b = 2 \pi$ while the size of $\chi$ is a free parameter.}  
The bulk reconstruction formula for a scalar field $\phi$ in the $R$ region has the same form as~\eqref{ebs} except that the integration over $q$ is replaced by a discrete sum. 
The corresponding boundary manifold, instead of being a Rindler patch of Minkowski spacetime, is given by $\RR \times S^1$. 
We take $\sM= \sM_R$, the algebra generated by single-trace operators in CFT$_R$. The corresponding algebra of bulk fields in the $R$ region is $\widetilde \sM_R$. 

%At finite $N$, the full (bounded) operator algebra of CFT$_R$ is type I. In the large $N$ limit, we are interested in 
%$\sM_R$,  the algebra  generated by single-trace operators of CFT$_R$. 

\begin{figure}[h]
\begin{centering}
	\includegraphics[width=2.5in]{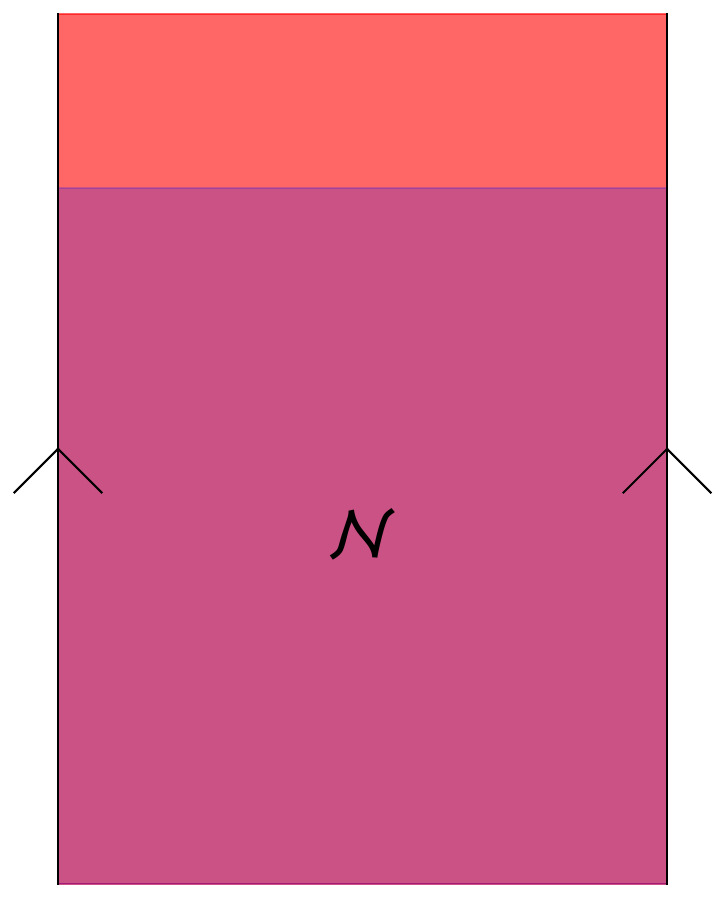} \qquad \includegraphics[width=2.5in]{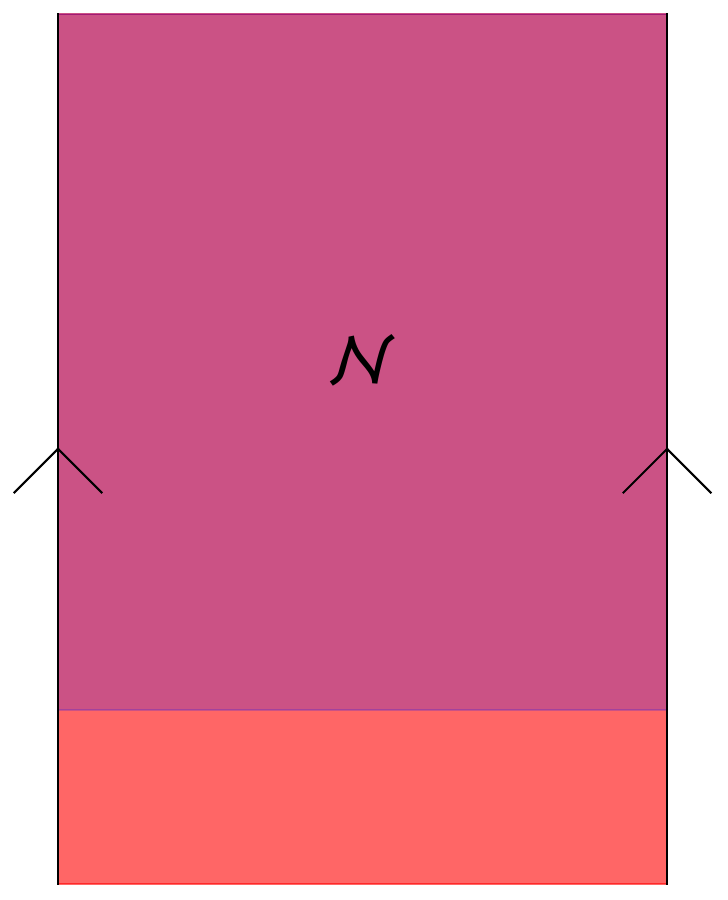}
\par\end{centering}
\caption{The left plot gives the boundary subregion that is used for the construction of a one-parameter group of unitaries that carries an observer from the $R$ region across the future horizon of the black hole into the $F$ region. The boundary is $ \RR \times S^1$ (vertical boundaries in the figure are identified). The relevant subalgebra is that generated by single-trace operators supported up to some maximum value of the boundary time (which we will take to be $\eta =0$). The right plot is the boundary subregion used to construct the unitaries to carry 
an observer from the $R$ region across the past horizon of the black hole. The relevant subalgebra is that generated by single-trace operators supported at boundary times greater than some minimum value (which we will again take to be $\eta =0$).  % \textcolor{red}{(I think that a cylinder could get confused with the diagram of $AdS_3$ that is sometimes drawn?)}
}
\label{fig:tShiftBdryBH}
\end{figure}

\begin{figure}[h]
\begin{centering}
	\includegraphics[width=2.5in]{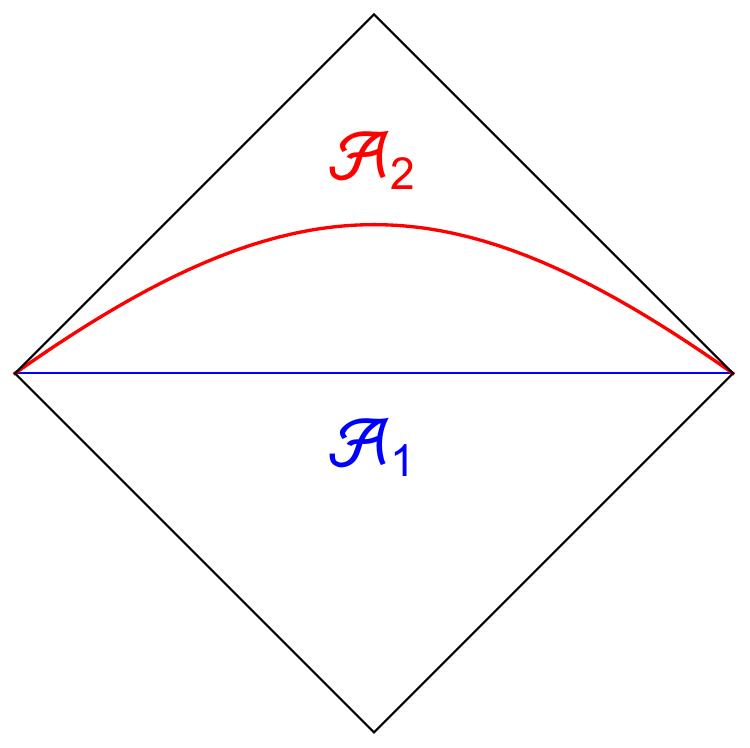}
\par\end{centering}
\caption{In a generalized free field theory, the algebras $\sA_1$ and $\sA_2$ of the two different slices shown are inequivalent, even though they share a causal diamond.
}
\label{fig:ineqBdryAlg}
\end{figure}

%Our goal is to demonstrate the emergence of horizon and interior $F, P$ regions of the black hole by explicitly constructing in-falling evolution operators.  

In this case the constructions of last section no longer apply since when $\chi$ is compact, a boundary null shift as in Fig.~\ref{fig:shiftedBdryRindlerWedge} is not well defined. We will instead take $\sN$ to be the operator algebra associated with the region indicated in the left plot of Fig.~\ref{fig:tShiftBdryBH}. 
To see that this is a sensible choice, it should be emphasized that generalized free fields do not satisfy any Heisenberg equations, and thus the algebras generated by them are not defined by causal diamonds. For example, the algebras associated with the two spacetime regions in Fig.~\ref{fig:ineqBdryAlg} are inequivalent, even though they share the same causal diamond.
The state $\ktfd$ is clearly separating with respect to $\sN$. While we do not have a rigorous mathematical proof, we will assume that it is also cyclic with respect to $\sN$. % \textcolor{red}{(I think that this is right approach for now. It should be possible to prove, but not as straightforward as originally thought)}

\iffalse
\sout{While we do not have a rigorous mathematical proof, we will assume that it is also} \textcolor{red}{Using a Reeh-Schlider-like argument, one may show that $\ktfd$ is also cyclic with respect to $\sN_R$~\cite{Borchers:1998ye}. \footnote{\textcolor{red}{BY mention on page 3 of their paper that, if the union of all time-translates of $\sN_R$ is dense in $\sM$, then there is a Reeh-Schlieder argument that $\ktfd$ will also be cyclic for $\sN_R$. Since the union of all time translates of our choice $\sN_R$ is $\sM$ I believe that this applies to our case so we can be sure that $\ktfd$ is separating for our choice of $\sN_R$. Perhaps this is worth understanding in more detail?}}}
\fi

\begin{figure}[h]
\begin{centering}
	\includegraphics[width=2.5in]{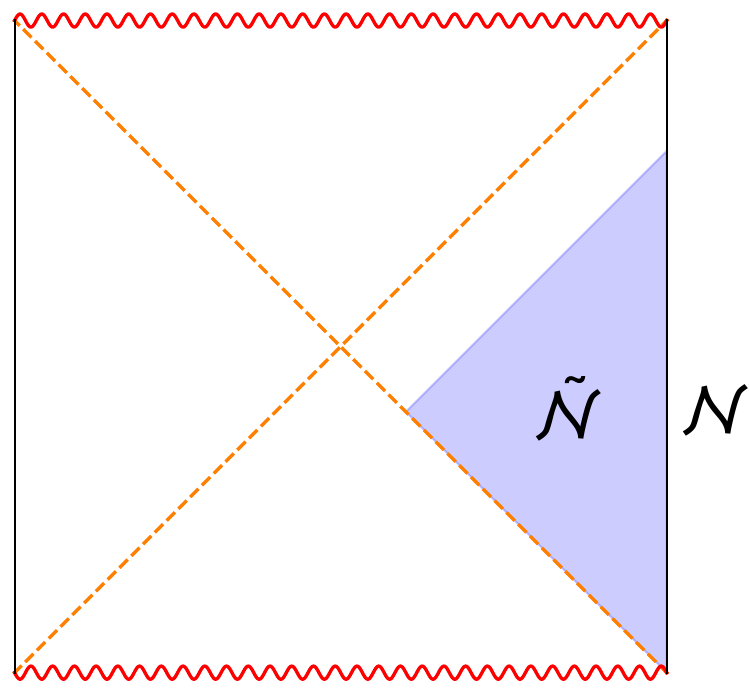} \qquad \includegraphics[width=2.5in]{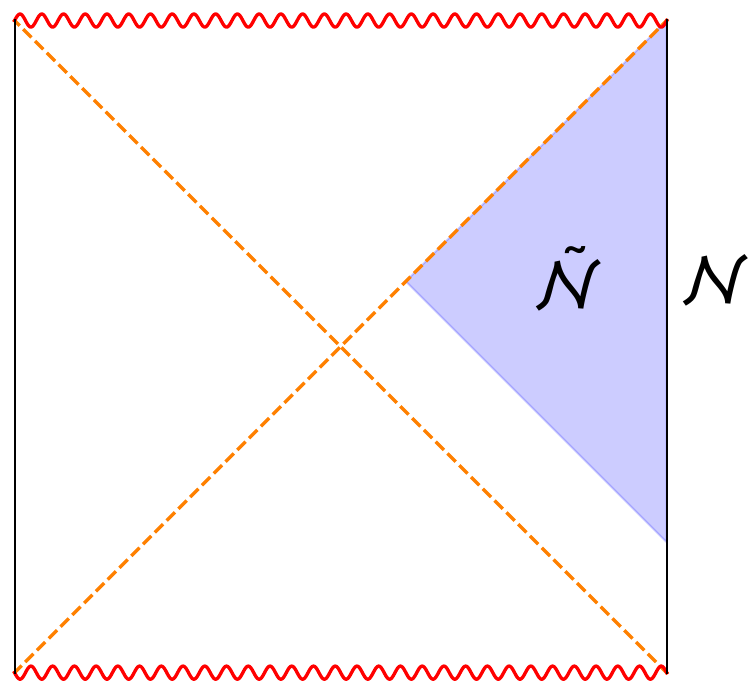}
\par\end{centering}
\caption{The respective proposed bulk duals for the boundary subregions %of operators 
indicated in Fig.~\ref{fig:tShiftBdryBH}. 
%supported below some maximum time is the blue shaded region.  
}
\label{fig:propEW}
\end{figure}

Finding $U(s)$ for this choice of $\sN$ is now difficult. We will find its explicit form by proposing a candidate for the dual of $\sN$ in the bulk.
%\footnote{\SL{In this section, we loosely use the term `entanglement wedge' to mean the bulk subregion whose associated operator algebra is equal to that of the corresponding boundary subalgebra.}} 
We propose that the operator algebra $\sN$ is dual to the algebra of bulk fields in the region $\widetilde \sN$ shown in Fig.~\ref{fig:propEW}.  
 This proposal is natural as this is the causal wedge associated with that part of the boundary, and we conjecture it is also the bulk subregion whose associated operator algebra is equal to $\sN$. We emphasize that the bulk dual here is not defined by an extremal surface prescription. 
We will  provide further support for this identification below.

Using the bulk theory it is then possible to explicitly construct $U(s)$. For a bulk scalar field~\eqref{ebs}
the corresponding matrix $C_{kk'}$ again has the form~\eqref{genSt} with phase $\ga$ given by 
\bega
e^{i \ga_k} = e^{i \de_k} {|\Ga (i \om)| \ov \Ga (i \om)} %= e^{- i \om \log 2} e^{i \th_k} {|\Ga (i \om)| \ov \Ga (i \om)}
= e^{- i \om \log 2}  {  \Ga (\bar q_- ) \Ga (\bar q_+) \ov |\Ga (\bar q_- ) \Ga (\bar q_+) |}
\label{eusm}
\end{gather} 
where $e^{i \de_k}$ is the phase shift for the scalar field at the horizon, and in the second equality we have given the explicit expression for the case of a BTZ black hole ($\bar q_\pm$ was defined earlier in~\eqref{qdef}). More explicitly, $\de_k$ can be read from the 
behavior of bulk mode function $v_k (w)$ near $w =1$
\be 
v_k (w) = {1 \ov \sqrt{2 |\om|}} \le( e^{- i \om \xi + i \delta_{k}} + e^{i
\om \xi - i \delta_{k}} \ri), \quad  w \to 1, \quad \xi = \ha \log{1-w \ov 1 + w} 
\ee
where $\xi$ is the tortoise coordinate. With the explicit form of $C_{kk'}$ we can then work out the action 
of $U(s)$ on $\phi_R (X)$. The resulting operator $\Phi (X, s)  \equiv U^\da (s) \phi_R (X) U (s)$ has the following properties: 

\ben 

\item It is not a local operator, but may be understood as $\phi$ smeared over a certain spacetime region. 
Writing $X = (\eta_0, \chi_0, w_0)$ in terms of Kruskal coordinates, $X = (U_0, V_0, \chi_0)$~(see Appendix~\ref{app:bh} for the transformation between the coordinates of~\eqref{Rindc} and the Kruskal coordinates), we find 
$\Phi (X; s)$ is supported only for $U < U_0 + s$.  In particular, for $s < s_0 \equiv  - U_0$, $\Phi (X; s) \in \widetilde \sM_R$, while for $s > s_0$, $a_k^{(L)}$  is now also involved. 

\item  For $X$ near the horizon, i.e. $-U_0 \ll 1$, 
$U(s)$ acts as a point-wise  $U$ translation, $\Phi (X; s) = \phi (X_s)$ with $X_s = (U_0 + s , V_0, \chi_0)$, to leading order in $-U_0$.

\begin{figure}[h]
\begin{centering}
	\includegraphics[width=2.5in]{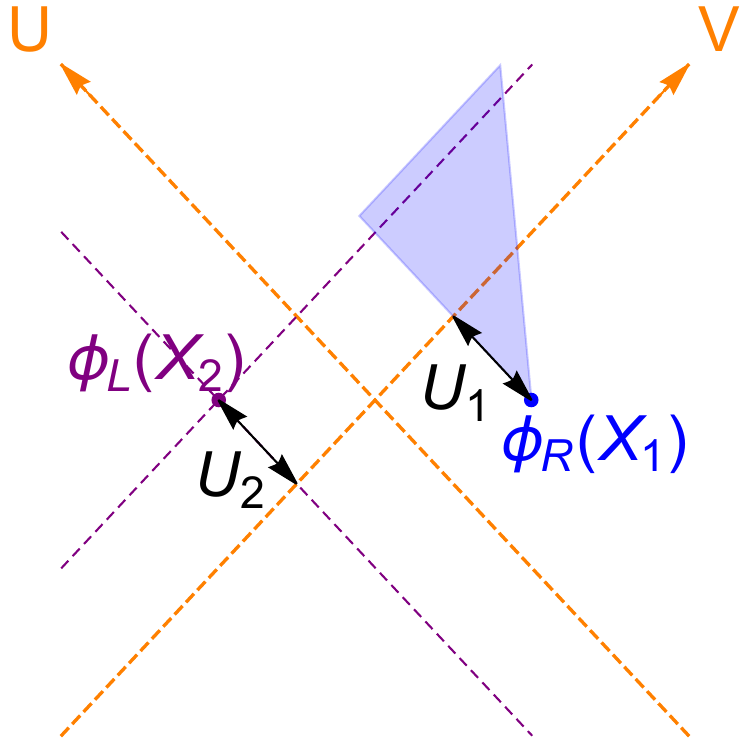}
\par\end{centering}
\caption{%Even thought the evolution is not point-wise, 
When a bulk field $\phi_R (X_1)$ with $X_1 \in R$ is transported by a null Kruskal coordinate distance $-U_1 + U_2$ (since $U_1 < 0$), it enters the lightcone of $\phi_L (X_2)$. The shaded region is a cartoon for the spread of $\Phi (X_1; s)$.  The orange dashed lines are event horizons, and the purple dashed lines give the light cones of $X_2$. The boundaries and singularities suppressed in the figure. 
%it is such that a right operator evolved by amount $u$ has support only at $U < U_1 + u$. This ensures that its commutator with a left operator is non-zero until we enter the lightcone of that operator, i.e. $u \geq -U_1 + U_2$.
}
\label{fig:commLightcone}
\end{figure}

\item For general points $X_1 \in R$ and $X_2 \in L$,  %= (\eta, w, \chi)$ 
we find that 
\be 
[ U^\da(s) \phi_R (X_1) U(s) ,  \phi_L (X_2)] =0
\ee
for $s < s_{12} \equiv -U_1 + U_2$, but the commutator becomes nonzero when $s >s_{12}$, precisely reproducing the casual structure expected from the black hole geometry.  See Fig.~\ref{fig:commLightcone}.
From the boundary theory perspective, this means that the two boundary systems are casually connectable in the sense of~\eqref{eja}. 

\item Acting on an operator $\sO_R$ at boundary point $(\eta_0,\chi_0)$ it is a nonlocal transformation with support only for $\eta < 
-  \log ( e^{-\eta_0} - s)$.  This agrees with~\eqref{bdrytshift}\footnote{Note the boundary time $\eta$ is related to modular time $t$ of~\eqref{bdrytshift} by  $\eta = \b t = 2 \pi t$.}, and provides a nontrivial consistency check of the identification of the shaded region $\widetilde \sN$ in Fig.~\ref{fig:propEW} as the bulk dual of the boundary subalgebra $\sN$.

\begin{figure}[h]
\begin{centering}
	\includegraphics[width=2.5in]{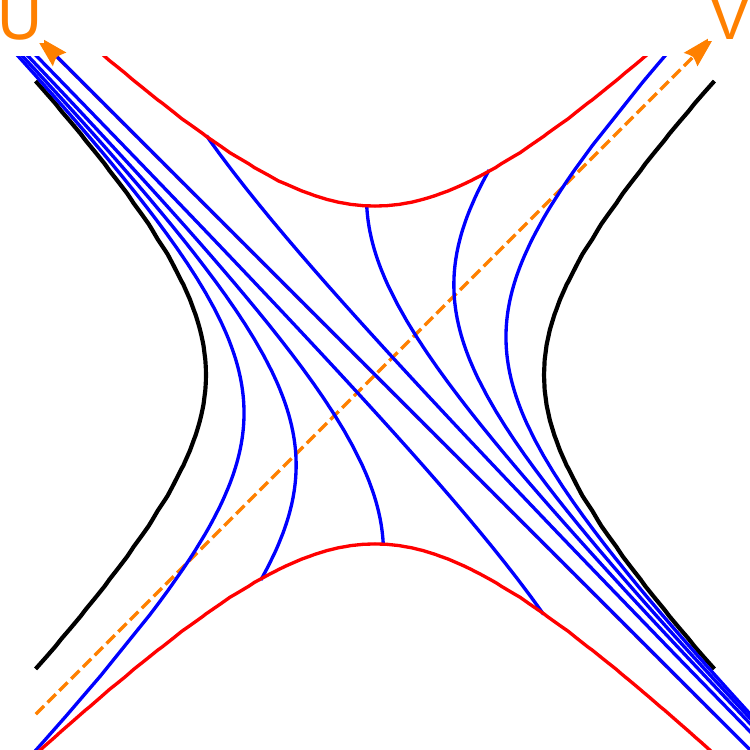} \qquad \includegraphics[width=2.5in]{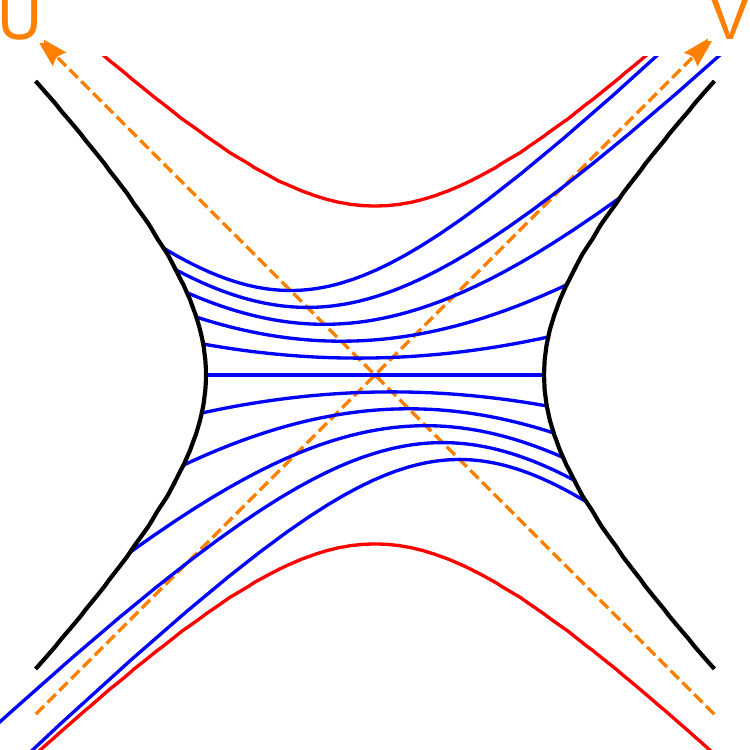}
\par\end{centering}
\caption{The left plot gives trajectories of~\eqref{heb}. The right plot gives constant $s$ surfaces evolved from the $\eta =0$ slice.  
The orange dashed lines are the event horizons, black solid lines are the boundaries, while the red solid lines are the singularities.   }
\label{fig:uEvolvedObserver}
\end{figure}

\item In the large $\Delta$ limit, with $q=0$ (i.e. if we dimensionally reduce both the bulk and boundary theories on the circle $\chi$), the transformation is  point-wise 
\be 
%U^\da (u) \phi_R (X) U(u) 
\Phi (X;s) =\lam_X  \phi (X_s), \quad \lam_X = \sqrt{1- s \sqrt{1-w_0^2} e^{\eta_0}}
\ee
with $X_s$ given by 
\bega 
e^{2\eta_s} %=-{V \ov 1- u V} {1 \ov U+u} %= {\sqrt{1-z \ov 1+z} e^{  t} \ov 1- u \sqrt{1-z \ov 1+z} e^{  t}}{1 \ov u -  \sqrt{1-z \ov 1+z} e^{- t }}
%= {\sqrt{1-z^2} e^t \ov 2u - u^2 \sqrt{1-z^2} e^t - \sqrt{1-z^2} e^{-t}}
=  {e^{2\eta_0} \ov 1- {2s e^{\eta_0} \ov \sqrt{1-w^2_0}}+ s^2  e^{2\eta_0} }, %\quad %t' = t -\ha \log \le(1- {2u e^{t} \ov \sqrt{1-z^2}}+ u^2  e^{2t} \ri)  \\
\quad w_s  %= {1-u V+ (U+u) V   \ov 1-u V-(U+u) V } 
%= {1+UV \ov 1- UV - 2 uV} 
= {w_0 \ov 1- s \sqrt{1-w^2_0} e^{\eta_0} } \ .
\label{zprim}
\end{gather} 
The above transformation can be expressed in terms of Kruskal coordinates as
\be \label{heb}
U_s = U_0+ s, \qquad V_s %= -{1 \ov \tilde U + s} 
= {V_0 \ov 1 - s V_0}  \ . %\qquad \tilde{U} = -\frac{1}{V}  \ .
\ee
The trajectories following from~\eqref{heb} are shown in Fig.~\ref{fig:uEvolvedObserver}. 

\een  

\begin{figure}[h]
\begin{centering}
	\includegraphics[width=2.5in]{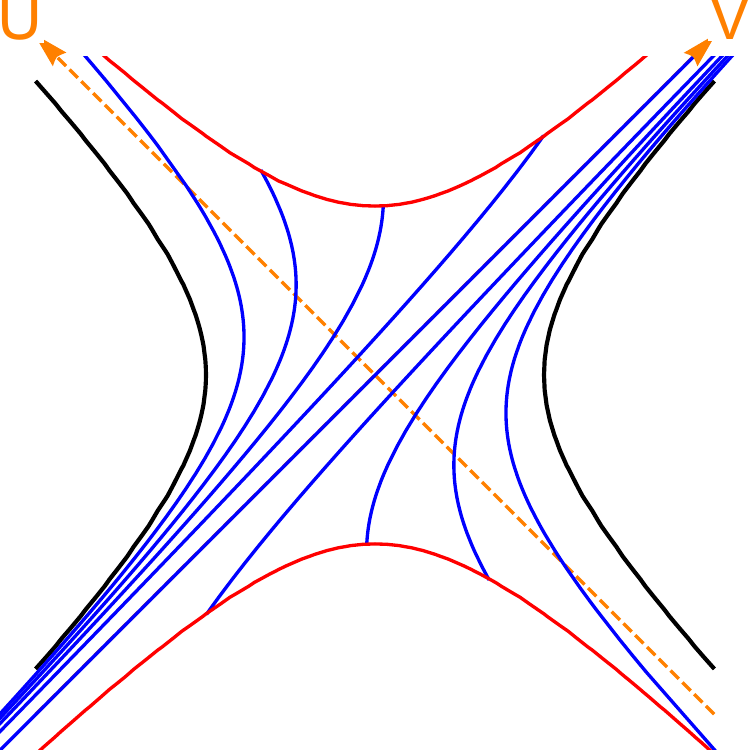} \qquad \includegraphics[width=2.5in]{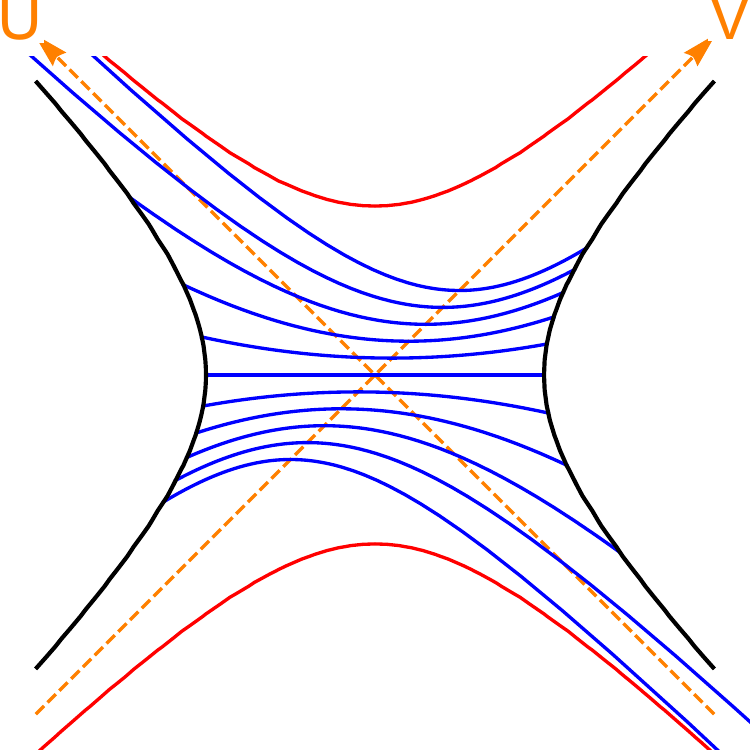}
\par\end{centering}
\caption{The counterparts of Fig.~\ref{fig:uEvolvedObserver} when using $\sN$ as in the right plot of Fig.~\ref{fig:tShiftBdryBH}.
}
\label{fig:vEvolvedObserver}
\end{figure}

By choosing $\sN$ to be the algebra associated with region in the right plot of Fig.~\ref{fig:tShiftBdryBH}, we can similarly construct 
unitary evolutions as above but with {the roles of Kruskal $U$ and $V$ swapped}. See Fig.~\ref{fig:vEvolvedObserver} for the corresponding flow trajectories.

\section{Discussion}

\subsection{Possible choices of $U(s)$ and more general states}

In Sec.~\ref{sec:btz} we discussed two possible choices of $\sN$ and the corresponding $U(s)$.
These are only the simplest choices. There are an infinite number of others. For example, for both plots in Fig.~\ref{fig:tShiftBdryBH} instead of letting the region describing $\sN$ be bounded by the $\eta =0$ slice on the boundary, we can choose a slice $\eta = f (\chi)$ where $\chi$ is the boundary spatial coordinate and $f$ an arbitrary periodic function. Alternatively, instead of taking the cyclic and separating vector $\ket{\Om}$ to be the ``vacuum'' $\ket{\Om_0}$ of the GNS Hilbert space dual to the bulk Hartle-Hawking vacuum we can also choose other $\ket{\Om}$. The simplest possibilities are obtained by acting on $\ket{\Om_0}$ unitaries from $\sM_R$ and $\sM_L$, i.e. 
\be \label{ienn}
\ket{\Om} = V_L W_R \ket{\Om_0}, \quad V_L \in \sM_L, \quad W_R \in \sM_R
\ee 
which results in a $U(s) = V_L W_R U_0 (s) W_R^\da V_L^\da$ with $U_0$ denoting the evolution operator corresponding to $\ket{\Om_0}$. 

Our discussion can also be generalized to more general entangled states of CFT$_R$ and CFT$_L$. A simple variant is to act on $\ktfd$ by a left unitary $U_L$ which does not change the reduced density matrix $\rho_\b$ of the CFT$_R$, i.e. 
\be 
\ket{\Psi} = U_L \ktfd \ .
\ee
The story depends on whether $\ket{\Psi}$ lies in the GNS Hilbert space built from $\ktfd$. If $\ket{\Psi}$ lies in the GNS Hilbert space, 
the bulk geometry is still described by the eternal black hole, now with some small excitations on the left due to insertion of $U_L$. The construction of $U(s)$ is the same as that for $\ktfd$. In particular, there are an infinite number of choices of $\ket{\Om}$ as in~\eqref{ienn}. When $\ket{\Psi}$ does not lie in the GNS Hilbert space, for example if $U_L$ changes the energy of the system by an amount which scales with $N$, {the story is different}. We need to work with the  GNS space $\sH_{\Psi}^{\rm GNS}$ associated with $\ket{\Psi}$, which does not overlap with that associated with $\ktfd$, and the corresponding representations $\sM_{L,R}$ of single-trace operator algebras are also different from those of associated with $\ktfd$.\footnote{The appearance of a different representation in this case is also required by the duality since the bulk geometry should also be modified.} In this case there is no simple relation between $U(s)$ for $\ket{\Psi}$ with those for $\ktfd$ as they act on different GNS Hilbert spaces. 

\subsection{Interpretation of the black hole singularity}

From a generic bulk point $X \in R$, the flow~\eqref{heb} reaches the future singularity for a finite value of $s$. 
Since we expect in general that the wave function of a bulk field should become singular at the singularity, the presence of the singularity in the black hole geometry should imply that the emergent evolution $U(s)$ breaks down at  finite values of $s$. 

It is instructive to contrast the nature of the $U(s)$ of Sec.~\ref{sec:btz} with those of Sec.~\ref{sec:rind}. 
In the discussion of Sec.~\ref{sec:rind}, while we also used generalized free fields, %which applies only the large $N$ limit, 
the evolution operator $U(s)$  from the choices of $\sN$ in Fig.~\ref{fig:shiftedBdryRindlerWedge} can be defined for the full theory at finite $N$. Thus $U(s)$ there should be well defined for $s \in (-\infty, +\infty)$. But the $U(s)$ of Sec.~\ref{sec:btz} only exists in the large $N$ limit, so the associated sharp horizon and interior (i.e. causal connectability from the boundary perspective) only exist in this limit. We can thus interpret the black hole singularity as a limitation on this emergent \causal; the connection of left and right observers cannot be extended indefinitely,\footnote{See~\cite{Nomura:2018kia} for a related perspective.} unlike the case of AdS Rindler.

\subsection{The nature of UV divergences in gravity and factorization of modular operator}

The entanglement entropy, $S_R$, of CFT$_R$ in $\ktfd$ is  given by the generalized entropy on the gravity side 
\be \label{ebg}
S_R = {A \ov 4 G_N} + \sS_R 
\ee
where $A$ is the horizon area and $\sS_R$ denotes entropy of matter fields in the $R$ region of the black hole geometry. 
There is also a relation between the corresponding modular operators~\cite{Jafferis:2015del}
\be \label{ebd}
H_R = {\hat A \ov 4 G_N} + K_R
\ee
where $H_R$ is the Hamiltonian of CFT$_R$, $K_R$ is the modular operator for the bulk field algebra $\widetilde \sM_R$ and $\hat A$ is the horizon area operator.  $K_R$ suffers from bulk UV divergences as does $\sS_R$.\footnote{Strictly speaking, $K_R$ cannot be mathematically defined due to UV divergences.}
But the left hand sides of~\eqref{ebg} and~\eqref{ebd} are well defined. For these expressions to make sense, the UV divergences of $K_R$ and $\sS_R$ must exactly be canceled by those in $G_N$ (understood as the bare coupling) to all orders in $G_N$ expansion. 

From the identification of $\sM_R$ and $\tilde \sM_R$, $K_R$ can be identified with the modular operator of $\sM_R$, and its divergences must then originate from the emergent type III$_1$ structure. This provides a different perspective on the bulk UV divergences and renormalization of the Newton constant $G_N$.\footnote{Recall that in the usual AdS/CFT dictionary, the bulk UV divergence is understood from the boundary theory as coming from a truncation of operators dual to stringy modes in the bulk.}
The divergence in $K_R$ is a reflection that, in a type III$_1$ algebra, the modular operator for $\sM_R$ cannot be factorized into those for the $R$ and $L$ systems. Since the algebra for the full CFT is type I, the corresponding modular operator is factorizable, and thus the area term in~\eqref{ebd} must ``restore'' the algebra from type III$_1$ to type I.\footnote{After this paper appeared, subsequent developments have suggested that the generalized entropy (up to an additive constant) can be obtained by deforming the algebra from type III$_1$ to type II$_{\infty}$~\cite{Witten:2021unn, Chandrasekaran:2022eqq}.}

\subsection{Some future directions} 

There are many more questions to be understood and we mention a few here. 
It is clearly of great interest to understand better the emergence of the type III$_1$ structure in the large $N$ limit,\footnote{The emergence of a type III$_1$ structure, and the associated symmetries  and continuous spectrum are also closely related to the discussion of~\cite{Goheer:2003tx} of incompatibility of an exact SL(2,R) symmetry with a finite number of states, and the factorization of Wilson line problem discussed in~\cite{Harlow:2015lma}.
} and what becomes of the in-falling evolution operators at finite $N$. In particular, it is important to understand more precisely the emergence of singularities in the boundary theory. The discussion here should also be generalizable to single-sided black holes including evaporating ones. We expect such constructions can shed new light on the information loss problem. 
We also expect that the manner in which an in-falling time emerges from the boundary theory here should teach us valuable lessons about holography for asymptotically flat and cosmological spacetimes. This should be especially helpful for understanding time in cosmological spacetimes including de Sitter.

\vspace{0.2in}   \centerline{\bf{Acknowledgements}} \vspace{0.2in}
We would like to thank Netta Engelhardt, Hao Geng, Daniel Harlow, Gary Horowitz, Daniel Jafferis, Lampros Lamprou, Juan Maldacena, Donald Marolf, Leonard Susskind, Aron Wall, and Edward Witten  for discussions, and Horacio Casini for communications. 
This work is supported by the Office of High Energy Physics of U.S. Department of Energy under grant Contract Number  DE-SC0012567 and and DE-SC0019127. 
SL acknowledges the support of the Natural Sciences and Engineering Research Council of Canada (NSERC).

\appendix 

\section{Kruskal coordinates for the BTZ spacetime}  \label{app:bh}  % and mode functions} 

For the BTZ metric~\eqref{Rindc}, 
the tortoise coordinate is given by
\be 
\xi = - \int {dw \ov 1-w^2} = \ha \log {1 - w \ov 1+w}   \ .
\ee
The Kruskal coordinates in the right exterior region are 
\bega\label{krubtz}
 U = -e^{\xi-\eta} =
- \sqrt{1-w \ov 1+w} e^{- \eta } , \quad
 V = e^{\xi+\eta} =
\sqrt{1-w \ov 1+w} e^{  \eta}, \\
- e^{2\xi} = UV = {w-1 \ov w+1} , \quad w = {1+ UV \ov 1- UV} , \quad
e^{2\eta } = - {V \ov U} \ .
\end{gather} 
The singularity lies at $UV = 1$ and the boundary at $UV =-1$. The
event horizons are at $U,V=0$.

\end{document}